\documentclass[aps,twocolumn,showpacs,superscriptaddress,floatfix]{revtex4-1}
\usepackage{graphicx,amsfonts,amssymb,amsmath,mathrsfs,hyperref}

\newif\ifhyper
\hypertrue
\ifhyper
\hypersetup{
   citecolor = {red},
   colorlinks = {true}, 
   linkcolor = {blue},
   urlcolor = {blue} 
}
\fi

\def\be{\begin{equation}}
\def\ee{\end{equation}}
\def\bea{\begin{eqnarray}}
\def\eea{\end{eqnarray}}

\def\tu{\triangle}
\def\td{\bigtriangledown}

\newcommand{\Zd}{\mathbb{Z}_2}
\newcommand{\A}{\mathcal{A}}

\newcommand{\BB}{\mathcal{B}}

\newcommand{\V}{\mathcal{V}}
\newcommand{\E}{\mathcal{E}}

\newcommand{\C}{\mathcal{C}}

 \newcommand{\ket}[1]{|#1\rangle}
 \newcommand{\bra}[1]{\langle #1|}
 
 \newcommand{\comutator}[2]{[ #1,#2 ]}
\newcommand{\anticomutator}[2]{\{ #1,#2 \}}

\begin{document}

\title{Topological spin liquids in the ruby lattice with anisotropic Kitaev interactions}

\author{Saeed S. Jahromi}
\email{jahromi@physics.sharif.edu}
\affiliation{Department of Physics, Sharif University of Technology, Tehran 14588-89694, Iran}

\author{Mehdi Kargarian}
\affiliation{Condensed Matter Theory Center and Joint Quantum Institute, Department of Physics, University of Maryland, College Park, MD 20742, USA}

\author{S. Farhad Masoudi}
\affiliation{Department of Physics, K.N. Toosi University of Technology, P.O. Box 15875-4416, Tehran, Iran}

\author{Abdollah Langari}
\email{langari@sharif.edu}
\affiliation{Department of Physics, Sharif University of Technology, Tehran 14588-89694, Iran}
\affiliation{Center of Excellence in Complex Systems and Condensed Matter, Sharif University of Technology, Tehran 14588-89694, Iran}

\begin{abstract}
The ruby lattice is a four-valent lattice interpolating between honeycomb and 
triangular lattices. In this work we investigate the topological spin-liquid 
phases of a spin Hamiltonian with Kitaev interactions on the ruby lattice using 
exact diagonalization and perturbative methods. The latter interactions combined 
with the structure of the lattice yield a model with $\Zd \times \Zd$ gauge 
symmetry. We mapped out the phase digram of the model and found gapped and 
gapless spin-liquid phases. While the low energy sector of the gapped phase 
corresponds to the well-known topological color code model on a honeycomb 
lattice, the low-energy sector of the gapless phases is described by an 
effective spin model with three-body interactions on a triangular lattice. A gap 
is opened in the spectrum in small magnetic fields, where we showed that the 
ground state has a finite topological entanglement entropy. We argue that the 
gapped phases could be possibly described by exotic excitations, 
and their corresponding spectrum is richer than the Ising phase of the Kitaev 
model. 
\end{abstract}
\maketitle

%
%
\section{Introduction}
\begin{figure*}[!htb]
\centerline{\includegraphics[width=18cm]{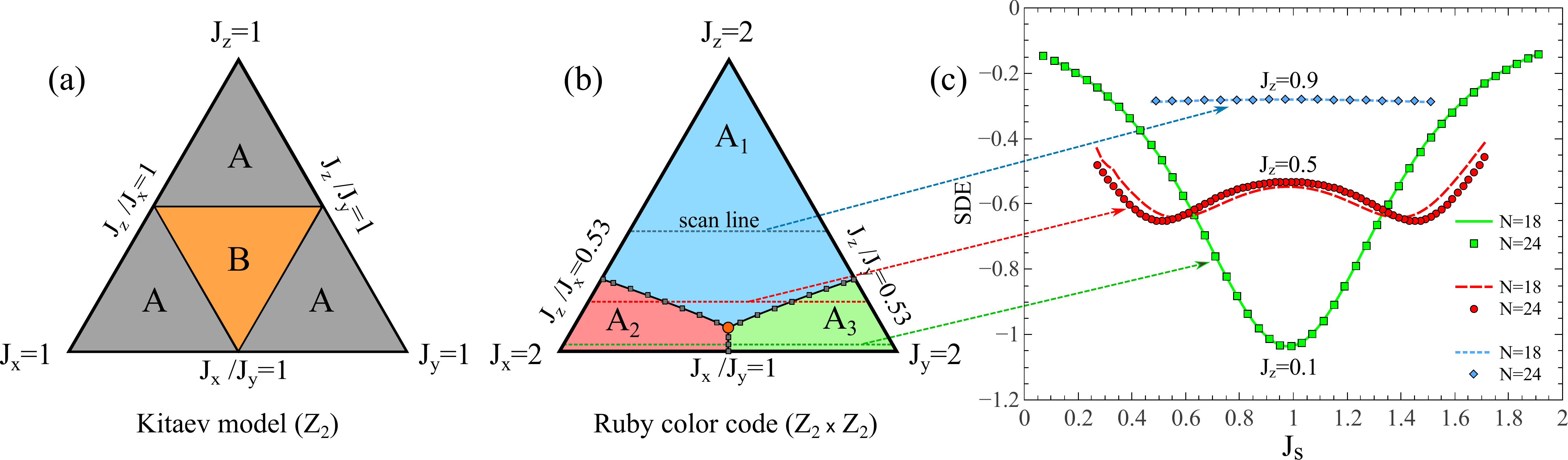}}
\caption{(Color online) Phase diagrams of (a) Kitaev model~\cite{Kitaev2006} in 
$J_x+J_y+J_z=1$ plane and (b) ruby color code model in (\ref{eq:two-body-TCC}) 
in the $J_x+J_y+J_z=2$ plane studied in this paper. The Kitaev model has the following 
phases: three symmetry-related gapped phases denoted by phase A, whose low energy description is given by toric code 
mode and the gapless phase B, which becomes a gapped phase with non-Abelian 
Ising anyons upon breaking time reversal symmetry. On the other hand the phase 
diagram of the RCC model consists of three phases labeled as $A_1, A_2, A_3$.
The $A_1$ phase is a topological gapped phase, and two symmetry-related gapless $A_2$ and $A_3$ phases. The phase boundaries are signaled by 
diverging, up to finite size effects, of the second derivative of the ground 
state energy per site, obtained by ED on periodic clusters with $18$ and $24$ 
sites. In (c) we show a few of them each corresponds to the line 
$J_{x}+J_{y}=2-J_{z}$ for a fixed $J_{z}$. We show this line by $J_{s}$.}
\label{Fig:phaseDiag}
\end{figure*}

Topological phases of matters have attracted a great deal of attention in recent years due to their novel properties such as topologically protected ground states \cite{Kitaev2003}, long-range entanglement \cite{Chen2010}
and emergent quasiparticles (QP) with fractional statistics, i.e. anyons \cite{Tsui1982, Wen1995, Wen1990, Wen2007, Kitaev2006}, which make them a suitable playground for topological quantum 
computation \cite{Nayak2008}. Our understanding of a topologically ordered phases in an exactly solvable spin model began with the toric code introduced by Kitaev \cite{Kitaev2003}. The ground state 
manifold is multiple degenerate depending on the genus of the space where the lattice is embedded, and the excitations carry Abelian statistics. However, the many-body nature 
of the spin interactions involving four-body terms in the underlying Hamiltonian makes its physical realization challenging. This problem was resolved by Kitaev in his 
seminal work \cite{Kitaev2006} by introducing a simple nearest-neighbor spin 
Hamiltonian on the honeycomb lattice. An exact solution based on Majorana 
representation of spins 
exists for the model yielding the Kitaev model two quantum spin liquid phases (see Fig.\ref{Fig:phaseDiag}(a) for a schematic representation) 
with $\Zd$ topological oder: a gapped phase which is continuously connected to the toric code and a gapless phase which can host non-Abelian Ising anyons when the 
Majorana fermions are gapped out by adding perturbations breaking the time-reversal symmetry (TRS). Although the toric model was first aimed at exotic excitations for quantum computations, recent experiments have unveiled the prospect of relevance of the Kitaev interactions in 
the highly anisotropic magnets on honeycomb lattices such as Na$_2$IrO$_3$, Li$_2$IrO$_3$ \cite{Singh2010,Reuther2011,Singh2012,Choi2012}, and $\alpha$-RuCl$_3$ \cite{Banerjee2016}.  
 
 \begin{figure}[t]
\centerline{\includegraphics[width=\columnwidth]{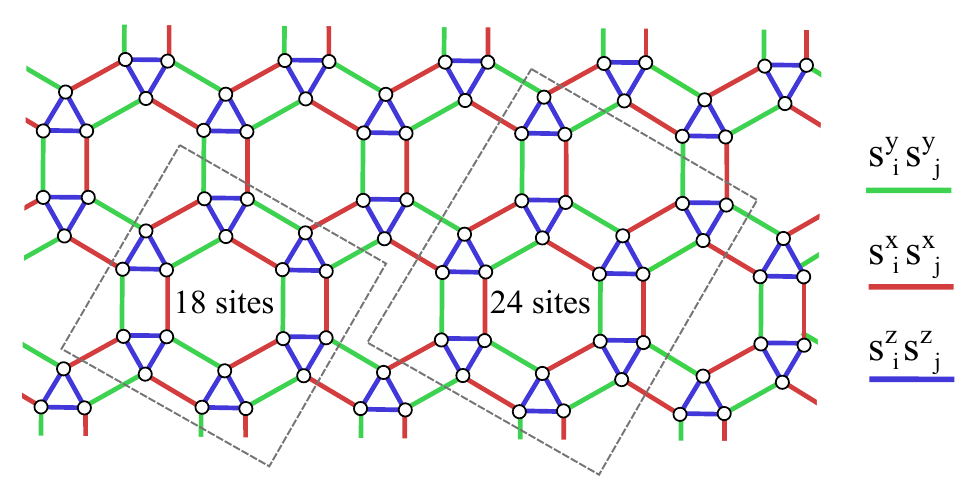}}
\caption{(Color online) ruby lattice $\Lambda$ of the two-body color code. Spin-1/2 particles are placed on the vertices of the lattice and the spin-spin interactions are denoted by
colored links. The rectangular dashed regions further represent the clusters used for exact diagonalization.}
\label{Fig:ruby}
\end{figure}
 
The discovery of such quantum magnets presaged the study of other models with anisotropic interactions on different lattices, including decorated honeycomb
\cite{Yao2007}, triangular \cite{Kargarian2012, Rousochatzakis2016}, spin 
ladder \cite{Karimipour2009, Langari2015} and ruby 
\cite{Bombin2009,Kargarian2010} lattices. 
The latter one, the ruby color code (RCC) shown in Fig.\ref{Fig:ruby}, is 
central to our work in this paper. The bismuth ions in layered materials such as 
Bi$_{14}$Rh$_3$I$_9$ form a bilayer ruby lattice \cite{Rasche2013,Pauly2015,Pauly2016} with 
interesting topological properties. We explore the phase diagram of a spin model 
with Kitaev interactions (\ref{eq:two-body-TCC}) on the ruby lattice in terms of 
exchange couplings ($J_{x},J_{y},J_{z}$) restricted to the plane 
$J_{x}+J_{y}+J_{z}=2J$.
Anisotropic interactions can in principle arise in transition metal compounds with strong spin-orbit couplings, which make the superexchange processes to be highly anisotropic and bond dependent \cite{Jackeli2009}.     

When $J_{z}\gg J_{x},J_{y}$, the low energy spectrum of the spin model (\ref{eq:two-body-TCC}) is gapped and the ground state manifold is topologically ordered \cite{Bombin2009,Kargarian2010}. The low energy sector is continuously connected to the so called topological color code (TCC) model first introduced by Bombin {\it et.al.} 
\cite{Bombin2006} to implement the Clifford group, transversally. 
In contrast to the Kitaev model, the topological order in TCC is associated with $\Zd \times \Zd$ gauge symmetry. This symmetry gives rise to emergence of highly interacting fermions with semionic
mutual statistics in the gapped phase~\cite{Bombin2009}.  

In this paper, we explore the full phase diagram 
of the RCC model, which to our best knowledge has not been explored so far. We 
contrast the phase diagram of $\Zd \times \Zd$ RCC  
with that of $\Zd$ Kitaev model; see Fig.\ref{Fig:phaseDiag}(a-b). 
We used finite-size exact diagonalization (ED) based on Lanczos algorithm on 
periodic clusters of different sizes to map out the full phase diagram of the 
RCC model from analysis of the ground state energy and its 
derivatives. Our results show that the phase digram contains one gapped and two gapless phases. The gapped phase corresponds to TCC on the honeycomb lattice as mentioned above. 
The gapless phases appear at the corner of phase diagram in the regime where either $J_{x}\gg J_{z},J_{y}$ or $J_{y}\gg J_{z},J_{x}$. This allows us to use degenerate perturbation 
theory (DPT) to derive the low-energy effective theory of the underlying phases.  
We find that the gapless phases are described by an effective Hamiltonian with three-body interactions on a triangular lattice. We argue that the latter phases could be possibly described by a rich structure of Ising anyons due to the underlying $\Zd\times\Zd$ gauge symmetry.

The paper is organized as follows: In Sec.\ref{Sec:Two-Body-TCC} we introduce the RCC model and review
some of the features of the model used in the paper. We present the phase diagram of the model in Sec.\ref{Sec:Phase-Diag} and characterize the 
underlying phases emerging in different coupling regime of the problem in Sec.\ref{Sec:Characterize}. A possible description of phases in terms of Ising anyons is discussed in Sec.\ref{Sec:Conclude}.

\section{The Model}
\label{Sec:Two-Body-TCC}
The RCC model \cite{Bombin2009} is a quantum spin system defined on a certain type of four-valent graphs, i.e. the ruby 
lattice $\Lambda$ shown in Fig.\ref{Fig:ruby}. The model is constructed by placing the spin-1/2 degrees of freedom on vertices of the lattice and
inducing two-body interactions of different types, distinguished by colored links, between nearest neighbors.
Hamiltonian of the RCC model is then defined as
\be
H=-\sum_{\alpha=x,y,z} J_{\alpha} \sum_{\alpha-\rm{links}} s_i^{\alpha}s_j^{\alpha} \quad,
\label{eq:two-body-TCC}
\ee
where the first sum runs over $\alpha$-links ($\alpha=x,y,z$) labeled by red ($\mathrm{r}$), green ($\mathrm{g}$) and blue ($\mathrm{b}$) colors, respectively, and the second sum runs over 
the two-body interactions acting on sites $i$ and $j$ of the $\alpha-\rm{links}$, and $s^{\alpha}$ stands for Pauli matrices. Here, we set $J_{\alpha}\!>\!0$.
The RCC model supports loop structures as well as string-net integrals of motions 
defined by connecting certain vertices and links of the lattice, underlying a $\Zd\times\Zd$ gauge symmetry \cite{Kargarian2010}.

In contrast to the Kitaev honeycomb model \cite{Kitaev2006}, the two-body color code on the ruby lattice is not exactly solvable because of the four-valence structure (four bonds are emanating from each site) of the  
lattice as opposed to the three-valence structure of the honeycomb lattice. 
Therefore, we resort to numerical techniques and approximation methods to map 
out the phase diagram of the Hamiltonian (\ref{eq:two-body-TCC}) in different 
coupling regimes ($J_x,J_y,J_z$). We restrict the exchange coupling to the 
$J_{x}+J_{y}+J_{z}=2J$. The coupling $J$ accounts for an overall energy scale, 
which we set to be unite $J=1$ throughout. 

\section{Phase diagram}
\label{Sec:Phase-Diag}
We apply the exact diagonalization technique
to the Hamiltonian (\ref{eq:two-body-TCC}) to capture the possible phases and phase transitions in different coupling regimes, by analyzing the ground state energy of the system and its derivatives. 
Our ED algorithm relies on the Lanczos method on the periodic clusters with $18$ and $24$ sites shown in Fig.\ref{Fig:ruby}. Matrix elements of the Hamiltonian are
further generated using the $s^z$ Pauli vector space and bit representation.

In order to capture the phase transitions of the model, we have calculated the second derivative of the ground state energy per-site (SDE) in the $J_x+J_y+J_z=2$ plane and detected the phase boundaries from the diverging behavior of the second derivatives of the energy as shown in Fig.\ref{Fig:phaseDiag}(c). Due to the finite size lattices, a real diverging is obscured. Thus, we take the location of minimum of SDE as a phase transition point. This might be a crude estimation of locating the phase transition, but we notice that on the paths in the phase diagram with possible phase transitions, e.g. the dashed red line in Fig.\ref{Fig:phaseDiag}(b), the behavior of SDE significantly differs from those paths with no phase transition. Moreover, moving from a lattice with 18 sites to a larger one with 24 sites, the minima in SDE become slightly deeper providing a strong evidence of phase transition. We elaborate on details on such transitions below. 

To find the phase boundary we look for the minima in SDE along paths 
corresponding to the intersection of the plane $J_x+J_y+J_z=2$ with a plane at 
fixed $J_z$ ($0\!<\!J_{z}\!<\!2$). We use $J_{s}$ accounting for a 
one-dimensional parameter space referring to the points lying on the intersection 
line. A few of such lines are shown as dashed lines in 
Fig.\ref{Fig:phaseDiag}(b), where we show the full phase diagram of the RCC 
model. We begin by setting $J_{z}=0.1$, the dashed green line. The corresponding 
SDE is plotted in Fig.\ref{Fig:phaseDiag}(c) with the same color to make the 
comparison with other SDE's easier. As seen, only one phase transition is 
signaled at ($J_x=0.95,J_{y}=0.95,J_{z}=0.1$). Increasing $J_{z}$ further, we 
didn't observe other phase transitions until a multi-critical point at 
$\mathbf{J}_c\approx(0.85,0.85,0.3)$ is reached, beyond which there are multiple 
phase transitions. We sum up this part by concluding that the region of the 
phase diagram with $0\!<\!J_{z}\!<\!0.3$ has two phases which we label as 
$A_{2}$ and $A_{3}$. The phase transition in this region occurs when 
$J_{x}=J_{y}$.           
 
Now we move the parameter line $J_{s}$ to go beyond the $\mathbf{J}_c\approx(0.85,0.85,0.3)$ point in the phase diagram, e.g. the dashed red and blue lines in Fig.\ref{Fig:phaseDiag}(b). Increasing $J_z\!>\!0.3$,
we observe that two distinct minima start to appear in the SDE curves. For 
instance let us consider the behavior of SDE as the parameter $J_{s}$ varies on 
the dashed red line in Fig.\ref{Fig:phaseDiag}(b). On this particular line 
$J_{z}=0.5$. We observed that two minima appeared in SDE. 
The first minimum signals a phase transition out of $A_2$ phase to another phase 
that we call it $A_1$, and the second minimum signals yet another phase 
transition from $A_1$ phase to $A_3$ phase. For $J_{z}$'s in the 
interval $0.3\!<\!J_{z}\!<\!0.69$ two minima appeared in SDE's, making phase 
boundaries between different phases marked by squares in 
Fig.\ref{Fig:phaseDiag}(b). Moving beyond the $J_{z}\!>\!0.69$, no phase 
transition appears, which shows that the RCC model is in the $A_{1}$ phase in 
this part of the phase diagram. Consider a parameter line $J_{s}$ corresponding 
to the dashed blue line $J_{z}=0.9$ on the phase diagram. The SDE plot is free 
of any minimum leading us to a conclusion that there is no more phase 
transition. 

\section{Low energy description of phases}
\label{Sec:Characterize}
The analysis presented in the preceding section yields a phase diagram with three distinct phases for the RCC model (\ref{eq:two-body-TCC}) within ED on finite clusters. According to the phase diagram Fig.\ref{Fig:phaseDiag}(b), each phase emerges when one of the couplings of the 
Hamiltonian is stronger than the two others. For example, the $A_1$ phase corresponds to the $J_{z}\gg J_{x},J_{y}$ coupling regime with strong interaction on 
the blue links, while the $A_2$ ($A_3$) phase emerges in the $J_x\!\gg\! J_y,J_z $ ($J_y\!\gg \!J_y,J_z $) coupling regime
with strong interaction on the red (green) links. 
Symmetry of the lattice structure further imposes that $A_2$ and $A_3$ phases to be equivalent, up to the interchange of the couplings ($J_y \leftrightarrow J_x $)
and colors of the red and green links. Here, we elaborate on the properties of the phases by focusing on each regime. 

\subsection{Topological Color Code: $A_1$ gapped phase}
\label{Sec:isolated-triangle-limit}
\begin{figure}[t]
\centerline{\includegraphics[width=\columnwidth]{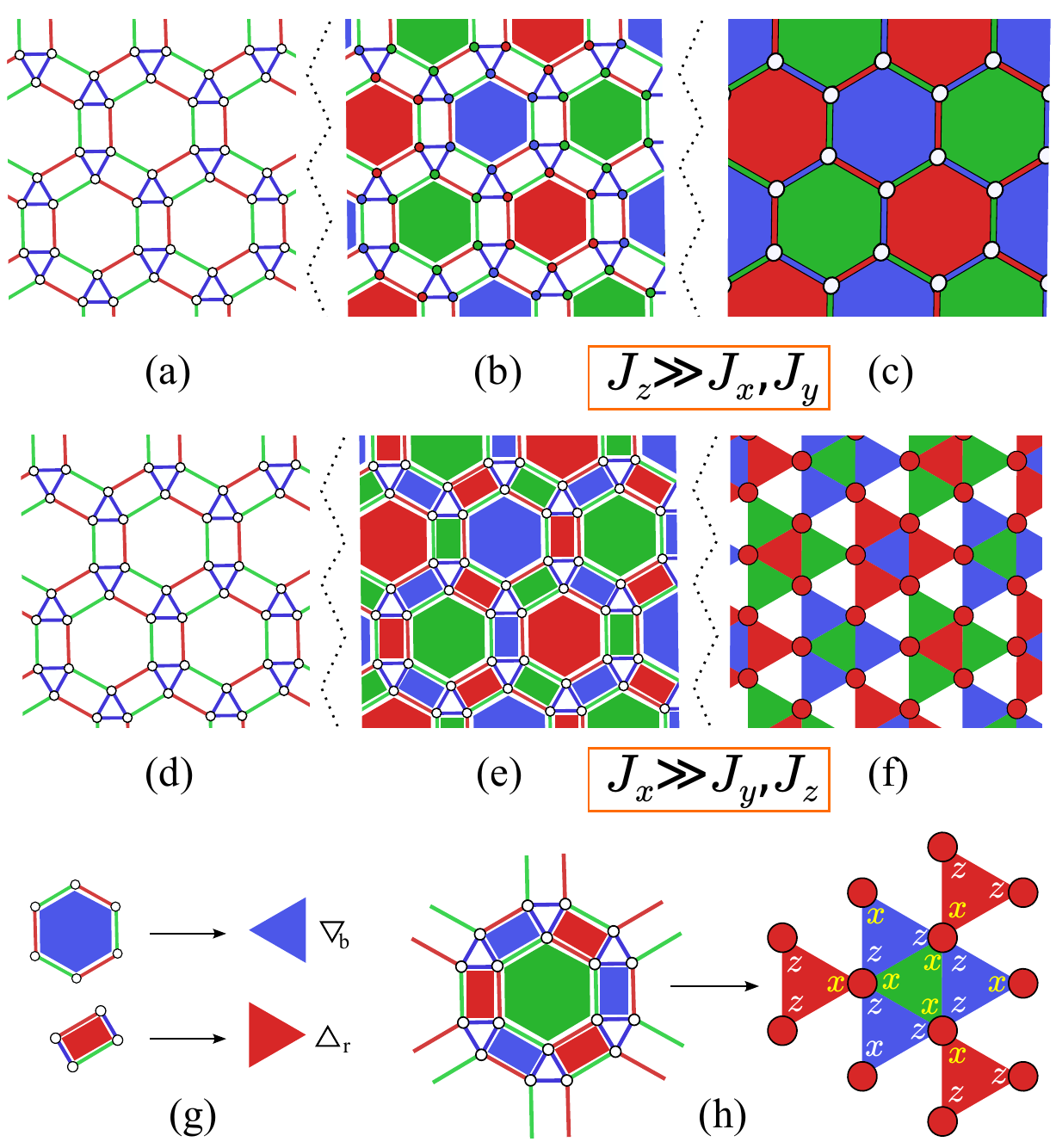}}
\caption{(Color online) The connection between original ruby lattice and effective honeycomb (a-c) and triangular (d-f) lattices corresponding to shrinking of blue triangles and red links to sites $v$. These limits, in the degenerate perturbation theory, correspond to $J_{z}\gg J_{x},J_{y}$ and $J_{x}\gg J_{z},J_{y}$ yielding effective Hamiltonians (\ref{Htcc}) and (\ref{eq:Low-Energy-H-DPT}), respectively. (g) Examples of the transformation of a blue hexagon and red rectangle to down triangle $\td_b$ and up triangle $\tu_r$, respectively. (h) A ruby plaquette and its corresponding triangular cluster in the effective language. The plaquette spin operators appearing in Hamiltonian (\ref{eq:Low-Energy-H-DPT}) are further shown by notations $x\equiv\tau^{x}$ and $z\equiv\tau^{z}$.}
\label{Fig:honey_tri}
\end{figure}

The $A_1$ phase arises in the particular regime of the couplings where $J_z\gg J_x,J_y$.
This regime of the problem has already been studied in detail in Ref.\cite{Bombin2009,Kargarian2010} and it has been shown that 
the low-energy physics of the Hamiltonian (\ref{eq:two-body-TCC}) in this limit is 
described by an effective topological color code model \cite{Bombin2006} on the honeycomb lattice; see Fig.\ref{Fig:honey_tri}(a-c). The low-energy description in this limit is given by a many-body Hamiltonian as follows
\bea \label{Htcc}
H_{\rm TCC}&=&-\sum_p (\tilde{J}_{z}Z_p+\tilde{J}_{x}X_p+\tilde{J}_{y}Y_p),
\eea
where sum runs over hexagonal plaquettes and the plaquette operators are product of Pauli matrices around a hexagon $Z_p=\prod_{i\in p} s_i^z$ and $X_p=\prod_{i\in p} s_i^x $ and $Y_p=\prod_{i\in p} s_i^y$. The coupling $\tilde{J}_{z}$ arises at 6th order of degenerate perturbation theory, while $\tilde{J}_{x}$ and $\tilde{J}_{y}$ arise at 9th order \cite{Kargarian2010}. The ground state of the model (\ref{Htcc}) is separated from the excited state by a gap. For a lattice with periodic boundary conditions defined on a torus with genus $g=1$ the ground state manifold is 16-fold degenerate resulting from the $\Zd\times\Zd$ gauge symmetry, as opposed to 4-fold degeneracy of the toric code with a $\Zd$ gauge group symmetry. Recently, a minimal TCC with seven qubits has been simulated in optical lattices being capable of detecting and correcting the errors \cite{Nigg2014}. The model has been the subject of several studies and many features of the model has already been revealed, ranging from 
error threshold \cite{Katzgraber2009}, robustness \cite{Jahromi2013a,Jahromi2013,Capponi2014}, entanglement properties \cite{Kargarian2008} and interesting quasiparticle excitations \cite{Jahromi2015}.

\subsection{Effective low-energy description of $A_2$, $A_3$ phases}
\label{Sec:isolated-dimer-limit}

Nature of the $A_2$ phase and low-energy physics of the Hamiltonian (\ref{eq:two-body-TCC}) in the limit $J_x\gg J_y,J_z $, 
to the best of our knowledge, is not a priori known. Similar limit for the Kitaev model on the honeycomb lattice has already been studied in Ref.\cite{Vidal2008,Dusuel2008}.
As discussed above, 
the phase $A_{3}$ arises in the limit $J_y\gg J_x,J_z$. Thus, we need to study 
one of them. We argue how the ruby lattice in the isolated-dimer limit is 
connected to a triangular lattice, and we setup a 
perturbative picture in the low-energy sector of the RCC model based on 
degenerate perturbation theory which is applied directly to the ($J_x\gg J_y,J_z 
$) limit of the Hamiltonian (\ref{eq:two-body-TCC}). 

The ruby lattice in the isolated-dimer limit is connected to a triangular lattice as shown in Fig.\ref{Fig:honey_tri}(d-f). This is best perceived by coloring the hexagons and the rectangles of the ruby lattice such that each rectangle connects two hexagons of the same color in the long direction and shares the same color with the hexagons. Following such coloring rules, the resulting colored ruby lattice is illustrated in Fig.\ref{Fig:honey_tri}(e). Next we replace the red links of the ruby lattice by red sites which shrink the hexagons and rectangles of the ruby lattice into {\it down} and {\it up} triangles labeled by $\td_c$ and $\tu_c$, respectively (see Fig.\ref{Fig:honey_tri}(g)). The subscript $c\in\{\mathrm{r},\mathrm{b},\mathrm{g}\}$ denotes the color of the reduced triangles.
The ruby lattice is then reduced to a triangular lattice labeled by $\tilde{\Lambda}$; see Fig.\ref{Fig:honey_tri}(f).

Next we use degenerate perturbation theory in the limit $J_x\gg J_y,J_z $ to 
derive an effective model on the triangular lattice $\tilde{\Lambda}$. Before that, let us for simplicity rotate the Hamiltonian (\ref{eq:two-body-TCC}) such that $(s^x,s^y,s^z)\rightarrow (s^z,s^y,-s^x)$
and then write it in the form $H=H_0+V$ where $H_0$ is the unperturbed diagonal part and $V$ is the perturbation represented by
\bea
H_0&=&-J_x\sum_{\rm r-link} s_i^z s_j^z, \label{eq:H0-2body}\\
V&=& -J_y\sum_{\rm g-link} s_i^y s_j^y -J_z\sum_{\rm b-link} s_i^x s_j^x, \label{eq:V-2body}
\eea
where $i,j$ denotes the nearest neighbors on the bonds of the ruby lattice. 
In the extreme case where $J_y,J_z=0$, 
the system is composed of isolated red dimers, where its ground state is given 
by $\ket{\uparrow\uparrow}$ and $\ket{\downarrow\downarrow}$ ferromagnetic 
states on the red links. Ground state of the system is therefore $2^{N_d}$-fold 
degenerate (where $N_d=N/2$ is the number of red dimers) with ground state 
energy $E_0=-N_dJ_x$. Excitations of the model correspond to antiferromagnetic 
red dimers that each cost $2J_x$ i.e. the first excited state of the system is 
$2N_d\times2^{N_d-1}$-fold degenerate and has a total energy $E_1=E_0+2J_x$. 
Effects of $J_y,J_z\neq 0$ interactions can further be studied perturbativally, 
around the strong $J_x$ couplings.

As we pointed out, the red dimers of the ruby lattice are equivalent to the vertices of the effective triangular lattice $\tilde{\Lambda}$. We therefore label each dimer by an index $v$ and define a projection operator on each dimer: 
\be\label{eq:projector}
P_{v}=\ket{\Uparrow}\bra{\uparrow\uparrow}+\ket{\Downarrow}\bra{\downarrow\downarrow},
\ee
where $\ket{\Uparrow}$ and $\ket{\Downarrow}$ are effective spin-1/2 on the 
vertex $v$ of lattice $\tilde{\Lambda}$. The ground state of $H_0$ is massively 
degenerate, and a weak perturbation $V$ lifts the degeneracy substantially. 
The low-energy sector then can be described by an effective Hamiltonian arising 
at the third order of perturbation. The details of the calculation are given in 
Appendix~\ref{appx:DPT}. The effective Hamiltonian reads as

\be
H_{\rm{eff}}^{(3)}=e_{0} + J_\td \sum_{\td\in\tilde{\Lambda}} \A_\td + J_\tu \sum_{\tu\in\tilde{\Lambda}} \BB_{\tu}
\label{eq:Low-Energy-H-DPT}
\ee
where 
\bea
\frac{e_{0}}{N}&=& -\frac{1}{2}-\frac{J_y^2}{2J_x}-\frac{J_z^2}{J_x} -\frac{J_y^3}{J_x^2},\\ 
J_\td&=&\frac{3J_y^3}{2J_x^2}, \;\;\;\;\; J_\tu=\frac{3J_yJ_z^2}{2J_x^2},\\
\A_\td&=&- \prod_{v \in \td} \tau_v^x , \label{eq:Ag} \\ 
\BB_{\tu}&=&- \prod_{v \in \tu} \tau_v^w, \quad \quad  
w=
\begin{cases}
x, & \text{ if } v \in \V \\
z, & \text{ if } v \in \E
\end{cases}, \label{eq:Bbg}
\eea
where $\tau_v^\alpha$ ($\alpha=x,z$) are the pseudo-Pauli operators acting on space spanned by $\ket{\Uparrow}$ and $\ket{\Downarrow}$ states.
On the triangular lattice $\tilde{\Lambda}$, each $\td_c$ triangle is surrounded by three $\tu_{\bar{c}}$ triangles which shares three edges with them
and is further connected to three $\tu_{\bar{\bar{c}}}$ triangles through its corners. Here the color changing bar operators are defined as
\bea 
\bar{\mathrm{r}}=\mathrm{g}, \quad \bar{\mathrm{g}}=\mathrm{b}, \quad \bar{\mathrm{b}}=\mathrm{r}.
\eea

Fig.\ref{Fig:honey_tri}(h) illustrates an example of a down triangle $\td_g$ which shares edges with the three neighboring up triangles $\tu_b$ and is connected to three other up triangles $\tu_r$ at its corners. Denoting the group of shared edges (vertices) by $\E$ ($\V$), structure of $\BB_{\tu}$ plaquette operator in Eq.~(\ref{eq:Bbg}) becomes clear.

Other orders of perturbation rather contribute to the ground state energy as an energy shift or produce terms that are always products of $\A_\td$ and $\BB_{\tu}$ plaquette operators. The overall low-energy effective theory of the RCC in the isolated-dimer limit is therefore given by (\ref{eq:Low-Energy-H-DPT}). Unlike the TCC model, which is exactly solvable, the anti-commutation of some plaquette operators appearing in (\ref{eq:Low-Energy-H-DPT}) obscures the exact solution. It is easy to see that $\anticomutator{\BB_{\tu}^c}{\BB_{\tu}^{c'}}=0$ when triangles share a site. 
Nevertheless, as shown in Appendix~\ref{appx:eff_string}, the model possess the $\Zd\times\Zd$ gauge symmetry. 

\begin{figure}
\centerline{\includegraphics[width=\columnwidth]{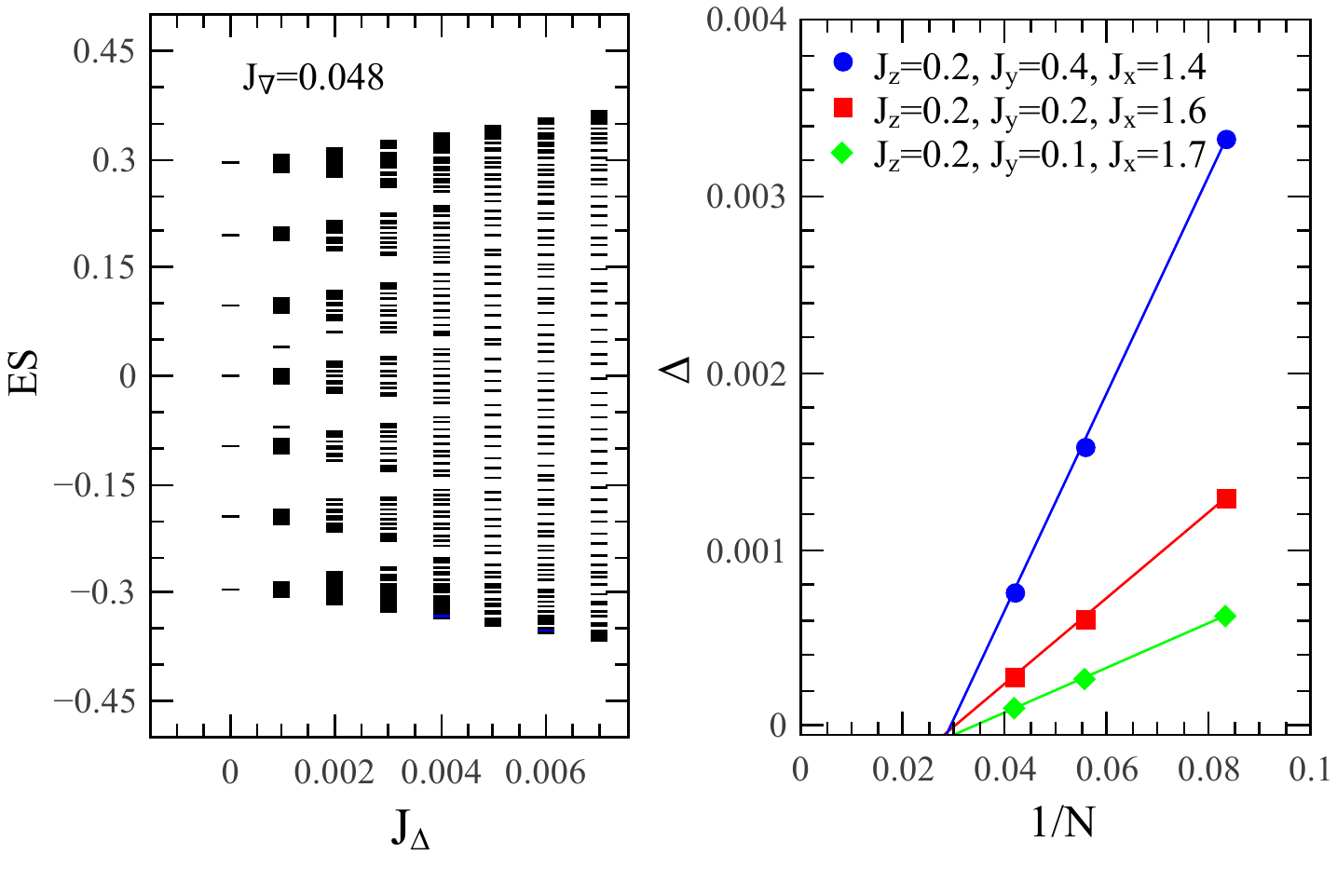}}
\caption{(Color online) (left) Energy spectrum of the effective Hamiltonian (\ref{eq:Low-Energy-H-DPT}) for $J_\td=0.048$ and varying  $J_\tu$ obtained by ED on 
periodic triangular clusters with $24$ sites. (right) Scaling of the energy gap $\Delta$ between the two lowest energy level for different couplings in the $A_2$ phase.
The gap approaches zero by increasing the system size indicating the gapless nature of the $A_2$ phase.}
\label{Fig:energy-spectrum}
\end{figure}

We, therefore, numerically explore the energy spectrum of (\ref{eq:Low-Energy-H-DPT}). In the extreme limit where $J_\tu=0$, the energy spectrum of $H_{\rm{eff}}$ is gapped as shown in Fig.\ref{Fig:energy-spectrum}. 
The most left pillar of the spectrum clearly shows the large gap between the degenerate ground states and the excited states of the 
the effective Hamiltonian (\ref{eq:Low-Energy-H-DPT}) at $J_\tu=0$. The energy spectrum of the effective Hamiltonian (\ref{eq:Low-Energy-H-DPT}) is studied by gradually increasing $J_\tu$. Surprisingly, even a very small $J_\tu$ would drastically change the energy spectrum and breaks the degeneracy of the ground state at 
$J_\tu=0$ coupling. Splitting of the energy levels at the bottom of the spectrum for different regimes of $J_\tu$ is clearly shown in Fig.~\ref{Fig:energy-spectrum}. 

In order to determine if the energy spectrum of the $A_2$ phase is gapped or gapless, we performed scaling over the energy gap between the two lowest 
eigenstates of the effective Hamiltonian (\ref{eq:Low-Energy-H-DPT}) in the ($J_\tu, J_\td \neq 0$) limits for different system sizes
on triangular lattice $\tilde{\Lambda}$ with $N=12, 18, 24$. The scaling was performed for different $J_\tu, J_\td$ couplings 
and $(J_x, J_y, J_z)$ were chosen such that to make sure we are deeply in the 
$A_2$ phase (see Fig.\ref{Fig:energy-spectrum}).
Our results certifies that the energy gap approaches zero by increasing the system size indicating the gapless nature of the $A_2$ phase. The same fact holds for the $A_3$ phase up to the interchange of $J_x$ and $J_y$ couplings.

\section{Summary and outlook}
\label{Sec:Conclude}
In this work we used numerical and perturbative methods to map out the phase 
diagram of the RCC model with $\Zd\times\Zd$ gauge symmetry, and the main 
results are summarized in Fig.\ref{Fig:phaseDiag}(b). We found three distinct phases are present in the 
phase diagram separated from each other by 
three phase boundaries met at a multi-critical point: (i) the $A_1$ is a gapped 
phase arising in the strong $J_z$ coupling, whose low-energy excitations are 
known to have Abelian statistics, and (ii) the $A_2$ and $A_3$ are two gapless 
phases arising in the regimes of couplings where either $J_x$ or $J_y$, 
respectively, is the strongest one. The low-energy description of latter phases are 
given by a three-body effective Hamiltonian (\ref{eq:Low-Energy-H-DPT}) on the 
triangular lattice. 

The latter phases are not continuously connected to a trivial paramagnetic phase 
in the presence of magnetic field; see Appendix \ref{appx:robust} for details. 
Indeed, we found there is a regime where the spectrum becomes gapped in the 
presence of a magnetic field. This behavior is not dissimilar to the magnetic 
field-induced gapped phase, the B phase in Fig.\ref{Fig:phaseDiag}(a), in the 
Kitaev honeycomb model \cite{Kitaev2006}. To determine whether the gapped phase 
is possibly a topologically ordered phase, we evaluated the topological 
entanglement entropy (TEE). The results are shown in Fig.\ref{Fig:gap-hz}. In 
contrast to the trivial polarized phase, which gives zero for TEE, the 
nonzero value of TEE in the gapped phase points to a distinct feature of this 
phase; the ground state could be topologically ordered.   

The exact determination of the nature of excitations in $A_2$ and $A_3$ phases is, however, rather elusive 
due to the lack of the exact solutions of RCC model (\ref{eq:two-body-TCC}) and three-body effective 
interactions (\ref{eq:Low-Energy-H-DPT}). However, we present a possible 
scenario below. We use an analogy with the Abelian and non-Abelian phases of the 
Kitaev model. The very low-energy description of the former is given by four 
super-selection sectors: the vacuum 1, the magnetic $m$ and electric $e$ 
particles, and the fermion $\epsilon=e\times m$. The latter phase is described 
by three super-selection sectors: the vacuum 1, the Ising anyons $\sigma$ and 
the fermion $\epsilon$. A connection between Abelian and non-Abelian Ising 
anyons has already been put forward \cite{Wootton2008, Bombin2010}. Especially, 
it is shown that the $\sigma$ particles can be identified form a superposition 
of strings of $m$ and $e$ anyons \cite{Wootton2008}: 
\bea |\sigma_{1}\sigma_{2};\pm\rangle=\frac{1}{\sqrt{2}}\left(|e_{1}e_{2}\rangle\pm|m_{1}m_{2}\rangle\right), \eea
where $e_{1}$ and $e_{2}$ are the end points of an open strings; the same holds 
for $m_{1}$ and $m_{2}$. The $\Zd\times\Zd$ Abelian gapped phase is basically 
two copies of the toric code model \cite{Kubica2015, Bombin2012, Bombin2014}. 
Thus, we expect the same construction can be used to identify the possible Ising 
anyons in RCC model. The low-energy sector of the Abelian phase is described by 
sixteen super-selection sectors \cite{Kargarian2010,Jahromi2015}: the vacuum 1, 
the anyons $\{e^{\mathrm{r}}, e^{\mathrm{b}}, e^{\mathrm{g}}, m^{\mathrm{r}}, 
m^{\mathrm{b}}, m^{\mathrm{g}}\}$, bosons $\{e^{\mathrm{r}}\times 
m^{\mathrm{r}}, e^{\mathrm{b}}\times m^{\mathrm{b}}, e^{\mathrm{g}}\times 
m^{\mathrm{g}}\}$, and the fermions $\{e^{\mathrm{r}}\times m^{\mathrm{b}}, 
e^{\mathrm{r}}\times m^{\mathrm{g}}, e^{\mathrm{b}}\times m^{\mathrm{r}}, 
e^{\mathrm{b}}\times m^{\mathrm{g}}, e^{\mathrm{g}}\times m^{\mathrm{r}}, 
e^{\mathrm{g}}\times m^{\mathrm{b}}\}$. Note that $e^{\mathrm{r}}\times 
e^{\mathrm{b}}\times e^{\mathrm{g}}=1$ and $m^{\mathrm{r}}\times 
m^{\mathrm{b}}\times m^{\mathrm{g}}=1$ due to $\Zd\times\Zd$ symmetry. 
Superposed the anyonic states, we obtain the Ising anyons as follows
\bea \label{Ising} |\sigma^{c}_{1}\sigma^{c}_{2};\pm\rangle=\frac{1}{\sqrt{2}}\left(|e^{\bar{c}}_{1}e^{\bar{c}}_{2}\rangle\pm|m^{\bar{\bar{c}}}_{1}m^{\bar{\bar{c}}}_{2}\rangle\right). \eea 

This suggest that two classes of colored Ising anyons, due to $\Zd\times\Zd$ symmetry, may arise in the gapless phases of the RCC model upon adding time-reversal breaking perturbations. Therefore, we conjecture that the Abelian $A_{1}$ phase undergoes a phase transition to $A_{2}$ and $A_{3}$ phases with colored Ising anyons $\sigma^{c}$. Viewed the topological color code as two coupled toric code models \cite{Kubica2015, Bombin2012, Bombin2014}, it suggests that the construction (\ref{Ising}) could be a spin analogue of coupled bilayer fractional quantum hall states with $\Zd$ symmetry. It is shown that for latter systems the condensation of Abelian anyons in the layers via a phase transition leads to rich structure for non-Abelian anyons such as $\mathrm{Ising}\times\Zd$ and $\mathrm{Ising}\times\mathrm{Ising}$ \cite{Vaezi2014a,Vaezi2014}. However, understanding the precise connection of this scenario to $A_{2}$ and $A_{3}$ phases requires more elaborative numerical studies, which can be a subject for future study.

\section{Acknowledgements}
The authors acknowledge Abolhassan Vaezi, Kai P. Schmidt, R. Haghshenas and H. Yarloo for helpful discussions. 
S.S.J and A.L acknowledge the support from the Iran National Science Foundation (INSF) under Grant NO. 93023859 and 
the Sharif University of Technology's Office of Vice President for Research.

\appendix

\section{Degenerate perturbation theory}
\label{appx:DPT}

In this section, we study the low-energy physics of the RCC Hamiltonian (\ref{eq:two-body-TCC}) in the
$J_x\gg J_y,J_z$ limit. Considering $H_0$ (\ref{eq:H0-2body}) as diagonal part of the RCC Hamiltonian, effect of the remaining parts
(\ref{eq:V-2body}) on $H_0$ can be studied as perturbation $V$. As we have pointed out in Sec.\ref{Sec:Characterize}, $H_0$ 
has a highly degenerate ground state subspace and a weak perturbation can lift the degeneracy substantially. We therefore apply the DPT technique
based on the projection operators and Green's function formalism \cite{Bergman2007} to extract the low-energy effective theory of the RCC model. Denoting the degenerate ground state subspace of the diagonal unperturbed part, $H_0$, by $\C$, the projection of any state $\ket{\Psi}$ to this subspace is given by $\ket{\Psi_0}=\mathcal{P}\ket{\Psi}$ where
\be
\mathcal{P}=\prod_v P_v.
\ee
and $P_v$, defined in (\ref{eq:projector}), is the projection from the 
$\ket{\uparrow\uparrow}$, $\ket{\downarrow\downarrow}$ physical qubits on sites $i,j$ of a red 
dimer on the ruby lattice $\Lambda$ to logical qubits on the vertex $v$ of the 
effective triangular lattice $\tilde{\Lambda}$. The projected state 
$\ket{\Psi_0}$ then satisfies the effective Schr\"{o}dinger equation in a 
perturbative level
\be
\left[ E_0+ \mathcal{P}V \sum_{n=0}^{\infty} \mathcal{G}^n \mathcal{P} \right]\ket{\Psi_0}=E\ket{\Psi_0}=H_{\rm eff}\ket{\Psi_0},
\ee
where $\mathcal{G}=\frac{1}{E-H_0}(1-\mathcal{P})V$.
The ground state energy $E$ can then be expanded in a series in perturbation parameters ($J_y,J_z$ in our case) within the degenerate manifold
\bea 
E=E^{(0)}_{0}+\sum_{k=1}^{\infty}E^{(k)}_{0},
\label{eq:Heff-DPT}
\eea
where $k$ is the order of perturbation.

According to the particular form of (\ref{eq:V-2body}), the perturbation $V$ 
would be a product of $s^x$ and $s^y$ Pauli operators, which act on 
different green and blue bonds of the ruby lattice in different orders of perturbation and take the ground state subspace to the excited state. However, there are particular
configurations of the bonds by acting on which, the ground state subspace is projected to itself i.e., preserves the ferromagnetic configurations of the dimers.

At zero order of perturbation, the effective Hamiltonian is denoted by $H_{\rm eff}^{(0)}=E_0^{(0)}$. The first order contribution is given by
\bea 
H_{\rm eff}^{(1)}=\mathcal{P}V\mathcal{P}. 
\label{eq:order1}
\eea
It is straightforward to check that the action of any two-body perturbation of the form $s_i^w s_j^w$ ($w=x,y$) on green and blue links, excites two red bonds to their antiferromagnetic 
configurations and takes the system out of its ground state manifold. Therefore, $\mathcal{P}V\mathcal{P}=0$ in the first order. In the second order of perturbation, the effective Hamiltonian reads
\bea 
H_{\rm eff}^{(2)}=\mathcal{P}VSV\mathcal{P},
\label{eq:order2}
\eea 
where $S=1/(E^{(0)}_{0}-H_{0})$. The second order consists of two $V$ terms and 
the only non-zero contribution which keeps the system in its ground state 
subspace originates
from those processes, wherein the the two $V$ terms double touch the blue and 
green bonds. In other words, the first $V$ excite two red dimers connected by a blue or green link to their excited states and the second $V$ returns them back to their original state. Therefore, in the second order the effective Hamiltonian acts trivially on the ground state
manifold and just shifts the ground state energy by
\bea 
H_{\rm eff}^{(2)}=-\frac{J_y^2}{2J_x} N-\frac{J_z^2}{J_x}N,
\label{eq:Heff2}
\eea
where $N$ is the number of lattice sites.

Order three is by far, the most interesting because the first non-trivial terms emerge at this order. The effective Hamiltonian at order three is given by
\bea
H_{\rm eff}^{(3)}=\mathcal{P}V \left(SV \right)^{2} \mathcal{P}. 
\label{eq:order3}
\eea

\begin{figure}[t]
\centerline{\includegraphics[width=\columnwidth]{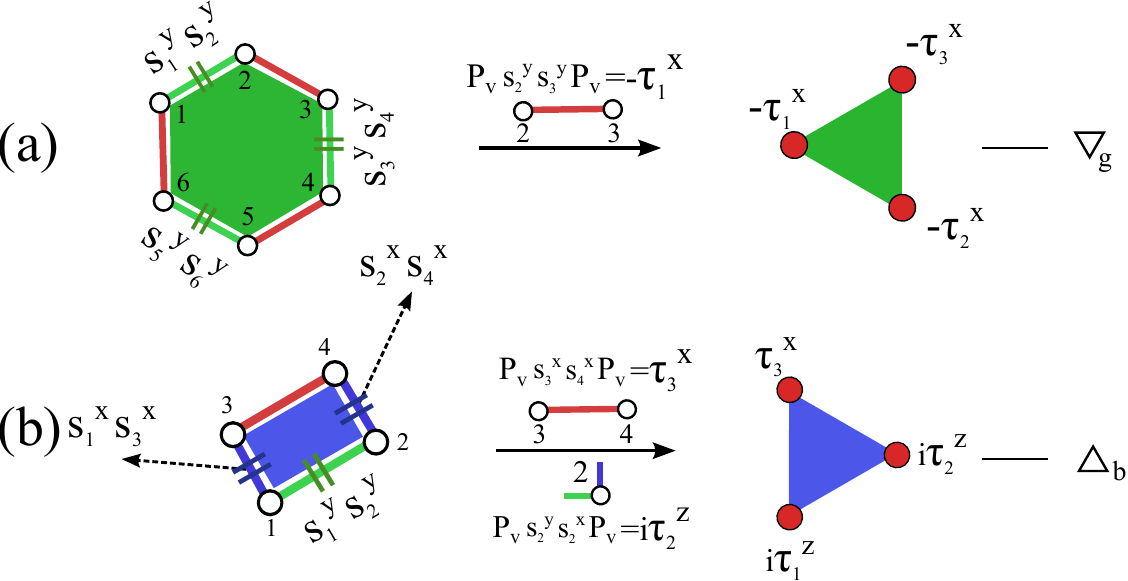}}
\caption{(Color online) 
The pictorial demonstration of the action of $H_{\rm eff}$ in order three of perturbation which shrinks the hexagons and 
rectangles of ruby lattice to the up and down triangles and encodes effective Pauli operators on the vertices $v$ of the triangles.
(a) Emergence of a $\td_g$ and (b) $\tu_b$ triangles and the corresponding plaquette operators. 
}
\label{Fig:o3-action}
\end{figure}

There are particular products of green and blue bonds, which map the subspace of 
system to itself and break the degeneracy of the ground state manifold
\be
\prod_{\langle ij \rangle \in {\rm g,b-link}} s_i^w s_j^w, \quad \quad w=x,y.
\label{eq:V3}
\ee
$H_{\rm eff}$ at the third order contains three $V$ terms which can act in two different ways with non-trivial outcome on the ruby lattice as depicted in Fig.\ref{Fig:o3-action}.
The first non-trivial term emerges from the product of three $s_i^{y} s_j^{y}$ on the green bonds of the inner hexagons of a plaquette on the ruby lattice (see Fig.\ref{Fig:o3-action}(a)). 
Such a product shrinks the inner hexagon of a ruby plaquette with color $c$ to a {\it down} triangle $\td_c$ and encode a logical $-\tau^x_v$
operator on each vertex $v$ of the $\td_c$ triangle. This can explicitly be seen from the following relation  
\be
P_v s_i^{y} s_j^{y} P_v=-\ket{\Uparrow}\bra{\Downarrow}-\ket{\Downarrow}\bra{\Uparrow} =-\tau_v^x,
\ee
where $P_v$ is the projector defined in (\ref{eq:projector}). Therefore, the expression (\ref{eq:V3}) at order three encodes 
the three-body plaquette operator $\A_\td$ (\ref{eq:Ag}) on the $\td_c$ triangles of the lattice $\tilde{\Lambda}$.

The next non-trivial term emerges from the action of two  $s_i^{x} s_j^{x}$ on blue links and one  $s_i^{y} s_j^{y}$ on the green bond of the 
rectangles of the ruby lattice as shown in Fig.\ref{Fig:o3-action}(b).
The action of $V$ then maps the the red dimer of the rectangle to an effective vertex 
and encode a logical $\tau_v^x$ operator on it. This process further encodes 
two logical $i\tau_v^z$ operators on the remaining vertices of the rectangle. As a result, the rectangle with color $c$ is reduced to 
an {\it up} triangle $\tu_c$. The projection can be best understood by noting that 
\bea
P_v s_i^{x} s_i^{y} P_v &=& P_v i s_i^{z} P_v=i\ket{\Uparrow}\bra{\Uparrow}-i\ket{\Downarrow}\bra{\Downarrow} = i\tau_v^z, \nonumber \\ \\ 
P_v s_i^{x} s_j^{x} P_v &=&\ket{\Uparrow}\bra{\Downarrow}+\ket{\Downarrow}\bra{\Uparrow}=\tau_v^x.
\eea
These operators all together encode the $\BB_{\tu}$ plaquette operator (\ref{eq:Bbg}) acting on the $\tu_c$ triangles the lattice $\tilde{\Lambda}$.

There is also another term at order three which arises from the action of three  $s_i^{x} s_j^{x}$ on the three connected bonds of a blue
triangle on the ruby lattice which has a trivial action on the ground state subspace and just shifts the ground state energy.
The low-energy spectrum of the system at order three of perturbation is then given by Hamiltonian (\ref{eq:Low-Energy-H-DPT}).

Similar to order two, the fourth order also shift the ground state energy, trivially. However, starting from order five, the non-trivial terms which break the degeneracy
again start to appear in the ground state manifold. One can check that these new terms are always the products of $\A_\td$ and $\BB_{\tu}$ plaquette operators.
The overall low-energy effective theory of the ruby color code model in the isolated-dimer limit is therefore given by (\ref{eq:Low-Energy-H-DPT}).

\section{String Operators and Integrals of Motion in $A_2$ phase}
\label{appx:eff_string}

In the previous section, we showed that the low energy physics of the $A_2$ phase is described by the effective Hamiltonian (\ref{eq:Low-Energy-H-DPT}).
With closer look at the model and the effective lattice $\tilde{\Lambda}$ as shown in Fig.\ref{Fig:honey_tri}-f, one can notice that the following commutation relations holds for the $\A_\td$ and $\BB_{\tu}$  plaquette operators
\bea
\comutator{\A_\td^c}{\A_\td^c}&=&\comutator{\A_\td^c}{\A_\td^{c'}}=0, \\ \label{eq:Ad-comute}
\comutator{\A_\td^c}{\BB_{\tu}^c}&=&\comutator{\A_\td^c}{\BB_{\tu}^{c'}}=0,\\
\comutator{\BB_{\tu}^c}{\BB_{\tu}^c}&=&0,\\
\comutator{\BB_{\tu}^c}{\BB_{\tu}^{c'}}&=&0 \quad \text{ if they share no sites},\\
\anticomutator{\BB_{\tu}^c}{\BB_{\tu}^{c'}}&=&0 \quad \text{ if they share a site}.\label{eq:Ad-Bu-anticomute}
\eea
Due to the latter anti-commutation relation, (\ref{eq:Ad-Bu-anticomute}), the effective Hamiltonian (\ref{eq:Low-Energy-H-DPT}) is not exactly solvable. However, the $\A_\td$ operator commutes with all terms of the $H_{\rm{eff}}$ and is therefore the integral of motion (IOM).
It is possible to show that the effective model further possess two other IOMs which can be produced either by the products of certain $\A_\td$ and $\BB_{\tu}$ operators or by going to high orders of perturbation, as will be shown subsequently. 

\begin{figure}
\centerline{\includegraphics[width=\columnwidth]{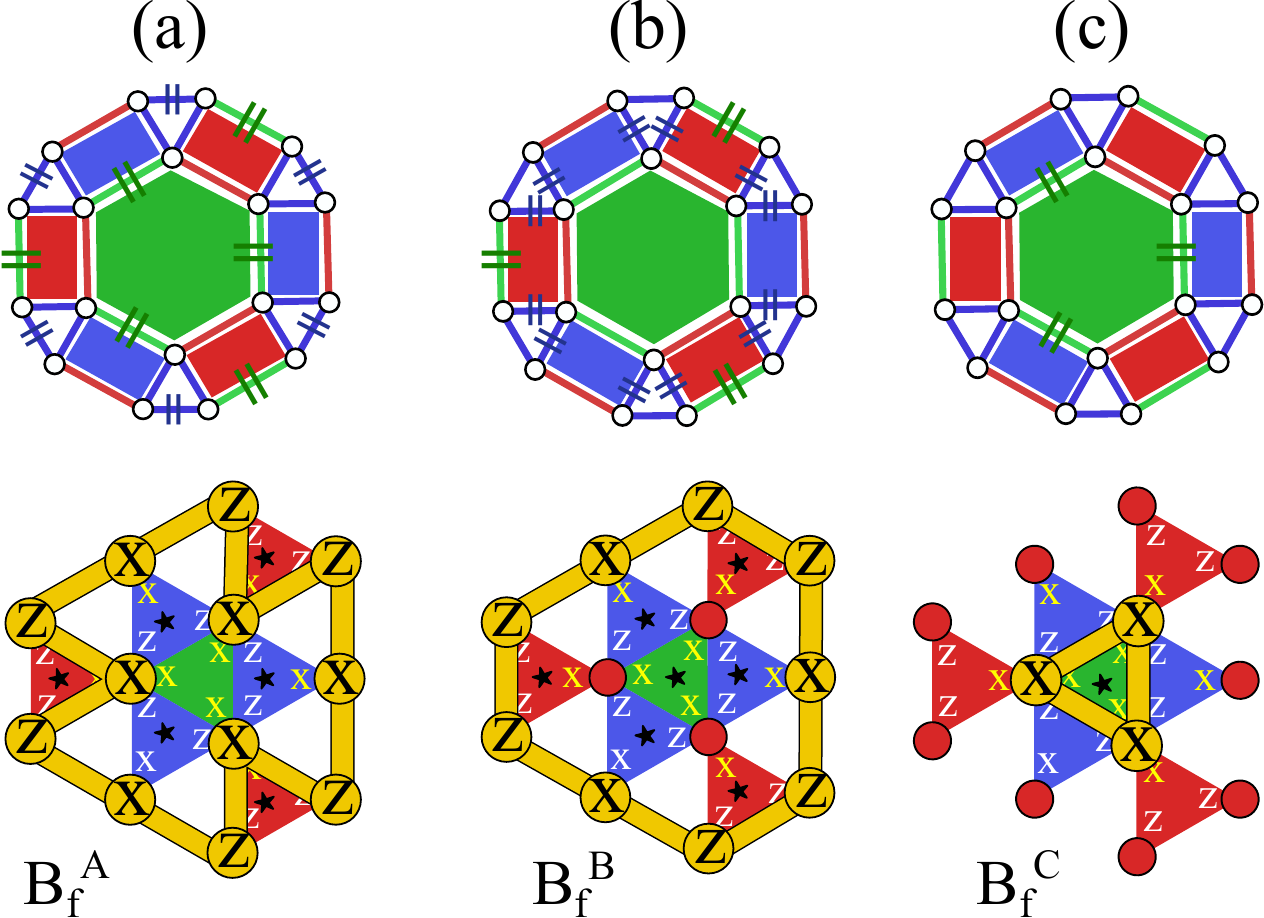}}
\caption{(Color online) (a-c)-top The links which are touched by $s_i^w s_j^w$ ($w=x,y$) interaction in orders $15, 12$ and $3$ with the net-effect of producing $B_f^A, B_f^B$ and $B_f^C$ 
elementary IOMs on the triangular lattice, respectively. (a-c)-bottom, The effective elementary IOMs on the triangular lattice. 
The black stars denotes the $\A_\td$ ,$\BB_{\tu}$ operators, which their product 
contribute in the structure of the elementary IOM operators.}
\label{Fig:eff-plaquettes}
\end{figure}

\begin{figure}[t]
\centerline{\includegraphics[width=\columnwidth]{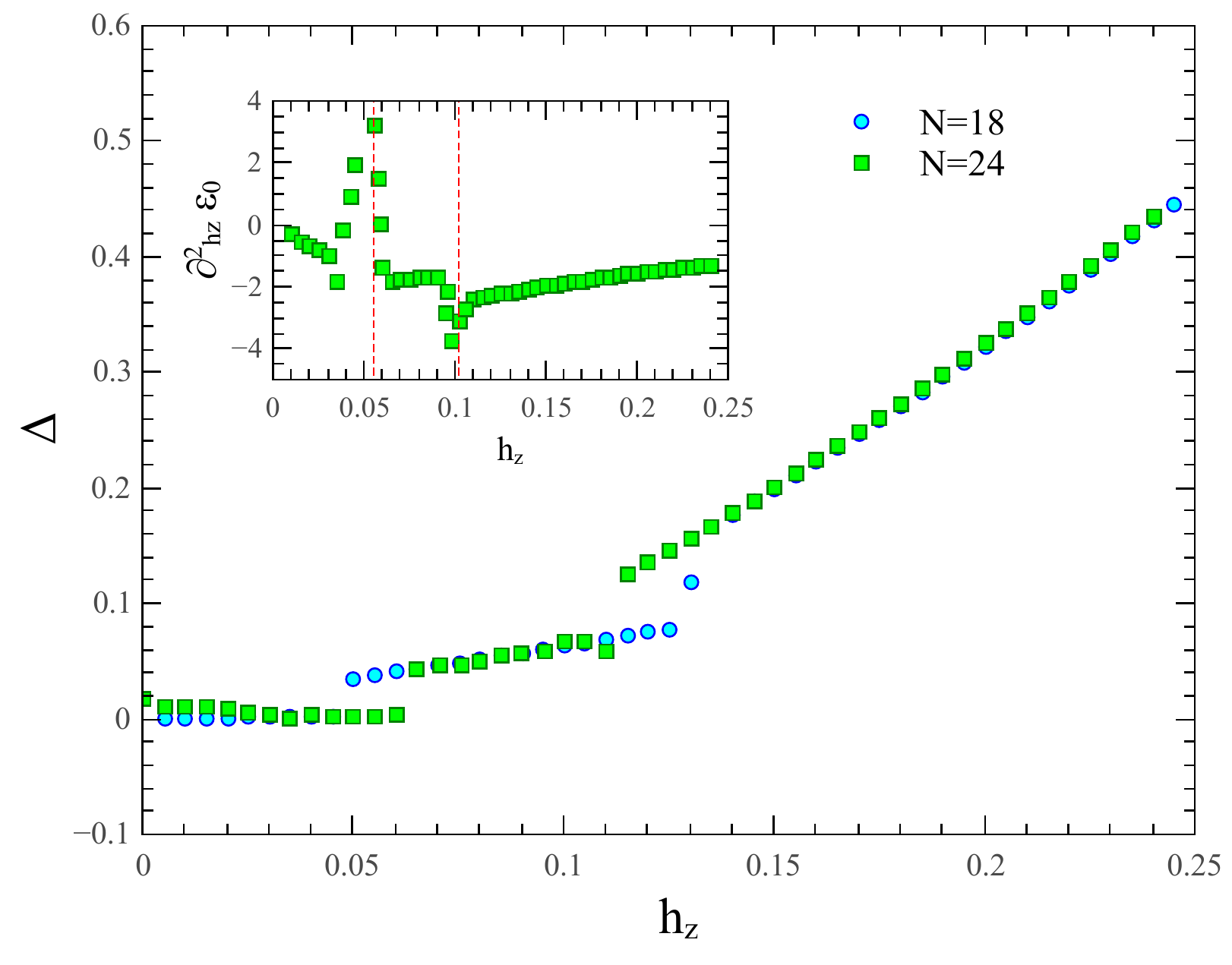}}
\centerline{\includegraphics[width=\columnwidth]{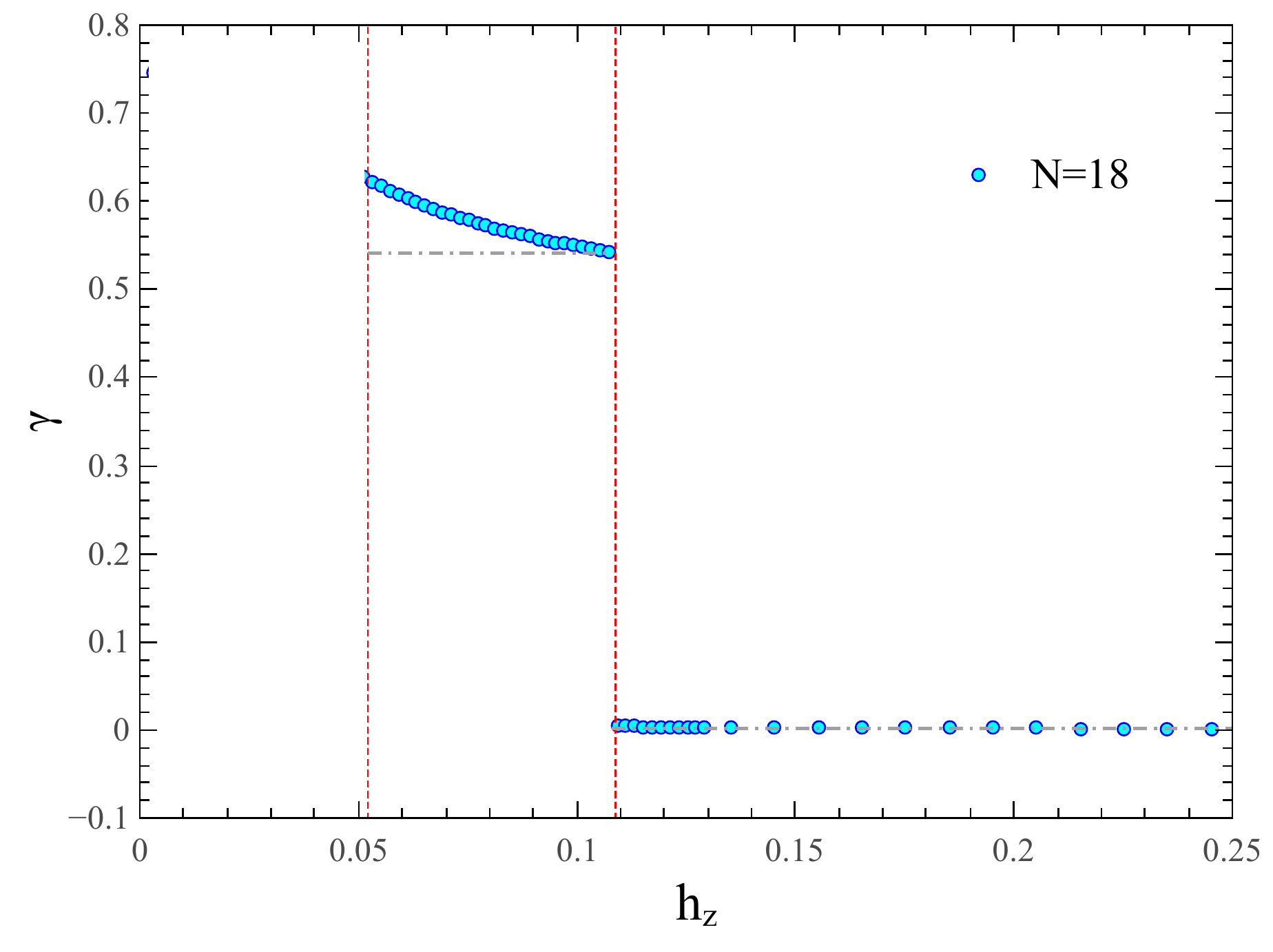}}
\caption{Upper panel: Energy gap of Hamiltonian (\ref{eq:two-body-TCC}) as a function of magnetic field in the $z$-direction obtained from ED on the ruby clusters with 18, 24 sites. The inset demonstrates the second derivative of ground state energy with respect to $h_z$. The red dashed lines in the inset
further demonstrate the location of the transition points. Lower panel: topological entanglement entropy (TEE) calculated for gapped phases. TEE drops to zero at $h_{z}\approx0.11$ where a phase transition between a topological phase and a trivial polarized phase occurs.}
\label{Fig:gap-hz}
\end{figure}

The second IOM of the model emerges at order twelve of perturbation. Similar to the procedure we envisaged in App.~\ref{appx:DPT}, there is a particular configuration for the action of two-body pertubations where six $s_i^{y} s_j^{y}$ and 
six $s_i^{x} s_j^{x}$ act on the green and blue links of the ruby plaquette as shown in Fig.\ref{Fig:eff-plaquettes}(a)-top. 
The action of $V$ terms then, projects the ground state to itself by reducing the red links of the lattice to effective vertices and encode the $B_f^A$
plaquette operator on the logical qubits as illustrated in Fig.~\ref{Fig:eff-plaquettes}(a)-bottom. 
The corresponding effective operator is given by 
\be
B_f^{A(B)}=-(+)\prod_{v} \tau_v^w, \quad \quad  
w=
\begin{cases}
x, & \text{ if } v \in \V \\
z, & \text{ if } v \in \E
\end{cases}
\label{eq:BfAB-DPT}
\ee
where $\V$ ($\E$) are the edges (vertices) shared by $\td_c$ and its surrounding $\tu_{\bar{c}}$ ($\tu_{\bar{\bar{c}}}$) triangles.

The last elementary IOM, $B_f^B$, is a closed string which appears at order fifteen from the action of 
three $s_i^{y} s_j^{y}$ and twelve $s_i^{x} s_j^{x}$ terms, respectively
on the green and blue links of the ruby plaquettes according to the convention shown in Fig.\ref{Fig:eff-plaquettes}(b)-top.
The $B_f^B$ plaquette operator is defined in (\ref{eq:BfAB-DPT}) and demonstrated in Fig.\ref{Fig:eff-plaquettes}(b)-bottom.

Defining $B_f^C=\A_\td$, it is immediately followed from the above relations that, locally, $B_f^A B_f^B=B_f^C$ 
and the local $\Zd \times \Zd$ symmetry of the RCC model is restored in the the $J_x\gg J_y,J_z$ limit. 
One can also check that the $B_f^A$ and $B_f^B$ plaquette operators can alternatively be constructed from the product of $\A_\td$ ,$\BB_{\tu}$ operators located inside $B_f^A$ and $B_f^B$. The contributing $\A_\td$ ,$\BB_{\tu}$ operators in the structure of each IOM are denoted by black stars in Fig.\ref{Fig:eff-plaquettes}.

On a triangular lattice with $N_t=N/2$ sites ($N$ is the number of sites on ruby lattice), there exist $N_t/3$ IOMs of each type ($A,B,C$) 
and the total number of $N_t$ elementary IOMs. The model therefore, possess $2^{N_t}$ independent IOMs.

\section{RCC in a magnetic field}
\label{appx:robust}
In this section, we study the stability of the gapless phases in the presence of a 
magnetic field in the $z$-direction, by analyzing the original RCC model (\ref{eq:two-body-TCC}) for $\mathbf{J}_{A_2}=(1.4, 0.4, 0.2)$ couplings.
The RCC Hamiltonian in the presence of the magnetic field is given by:
\be
H'=-\sum_{\alpha=x,y,z} J_{\alpha} \sum_{\alpha-\rm{links}} 
s_i^{\alpha}s_j^{\alpha}-h_z \sum_i s_i^z \quad.
\label{eq:two-body-TCC-hz}
\ee

In the extreme case where $h_z=0$, the system is in the $A_2$ phase, which 
according to our numerical results (see Sec.\ref{Sec:Characterize}) is a gapless 
phase. However, in the high magnetic field limit where $J_\alpha=0$, ($\alpha=x,y,z$), the 
Pauli spins are all aligned in the field direction and the ground state 
of the system is given by a polarized phase in the $z$-direction.
The low-lying excitations over this polarized ground states are denoted by single spin flips each with $2h_z$ energy cost. The system is 
therefore gapped. 
When all couplings are non-zero, at least a phase transition between the 
$A_2$ gapless phase and the gapped polarized phase of the high-field limit is 
expected. Other intermediate phases may also emerge in between. In order to 
capture the possible phase transitions, we calculated the energy gap of the 
system as a function of $h_z$, as well as the ground state energy of the system 
and its derivatives.

The energy gap for different values of magnetic field $h_z$ is shown in the 
upper panel of Fig.\ref{Fig:gap-hz}. The results show that the $A_2$ gapless 
phase is stable up to a finite field at $h_z^{c_1}\approx0.05$ where a phase 
transition occurs to an intermediate gapped phase. This latter phase is not 
continuously connected to a trivial polarized phase arising at high magnetic 
field. Indeed, a second phase transition to a polarized phase occurs at 
$h_z^{c_2}\approx0.11$. In the inset of this plot we also show the SDE with 
respect to $h_z$; it clearly shows two phase transitions signaled by divergences 
of SDE.

The intermediate gapped phase could be a topological phase distinct from a trivial paramagnetic phase. To explore the topological properties, a natural way would be to evaluate the topological entanglement entropy (TEE) of gapped phases. Given a normalized wave-function $\ket{\phi}$ and a partition of the system into subsystems $A$ and $B$, the reduced density matrix of subsystem $A$ is given by $\rho_A={\rm Tr}_B \ket{\phi}\bra{\phi}$. The von Neumann entropy $S=-{\rm Tr} (\rho_A \log_2 \rho_A)$ measure the entanglement between two subsystems. For a 2D topologically ordered gapped phase, the latter quantity assumes an area law scaling as $S=\alpha L-\gamma+ \mathcal{O} (1/L)$ \cite{Kitaev2006a,Levin2006},
where $L$ is the length of the region $A$ with smooth boundary. In this expression the first term arises from the non-universal and local contribution of the entanglement entropy. The second term $\gamma$, however, is a universal
constant being a signature of a topologically ordered phase. Distinctive feature of a topological phase is signaled by nonzero $\gamma$. We evaluated $\gamma$ as function of magnetic field $h_{z}$ for gapped phases. The results are shown in the lower panel in Fig.\ref{Fig:gap-hz}. It shows that the TEE is nonzero in the intermediate gapped phase and drops to zero at the phase transition to the trivial polarized phase. 

Let us note that the determined phase boundary suffers from the finite size effects and more accurate results might be obtained by performing the calculations on larger lattice sizes using more powerful numerical arsenals. However, the non-zero TEE for small magnetic fields and its transition
to zero TEE for $h_z>0.11$ is a clear signature of two topologically distinct phases.

%

 \end{document}